\documentclass[letter]{aa} % for the letters 

\usepackage{epsfig}     
\usepackage{graphicx,color}     
\usepackage{amssymb}            
\usepackage{url}                
\usepackage{amsmath}            
\usepackage{rotating}                   
\usepackage{float}                      
\usepackage{textcomp}           
\usepackage{epstopdf}
\usepackage{dcolumn}
\usepackage{times}
\usepackage{tabularx}
\usepackage{hyperref}
\usepackage[normalem]{ulem}
\hypersetup{
    colorlinks,
    citecolor=blue,
    filecolor=blue,
    linkcolor=blue,
    urlcolor=blue,
    menucolor=black}
\usepackage{soul} 
\usepackage[english]{babel}
\usepackage{booktabs}
\usepackage{gensymb}
\usepackage{subfig}
\usepackage{caption}
\usepackage{academicons}
\usepackage{orcidlink}

\newcommand{\hrieuv}{HRI\textsubscript{EUV}\xspace}

\begin{document}

\title{Evolution of dynamic fibrils from the cooler chromosphere to the hotter corona}

\author{Sudip~Mandal\inst{1}\orcidlink{0000-0002-7762-5629} ,
Hardi~Peter\inst{1},
Lakshmi~Pradeep~Chitta\inst{1}, 
Sami~K.~Solanki\inst{1,2},
Regina~Aznar~Cuadrado\inst{1}\orcidlink{0000-0003-1294-1257}, 
Udo~Sch\"{u}hle\inst{1}\orcidlink{0000-0001-6060-9078},
Luca~Teriaca\inst{1}, 
Juan~Mart\'inez-Sykora\inst{9},
David~Berghmans\inst{5}, 
Fr\'{e}d\'{e}ric~Auch\`{e}re\inst{6}, 
Susanna~Parenti\inst{6}\orcidlink{0000-0003-1438-1310},
Andrei~N.~Zhukov\inst{5,7}\orcidlink{0000-0002-2542-9810},
\'{E}ric~Buchlin\inst{6}\orcidlink{0000-0003-4290-1897},
Cis~Verbeeck\inst{5},
Emil~Kraaikamp\inst{5},
Luciano~Rodriguez\inst{5},
David~M.~Long\inst{10}\orcidlink{0000-0003-3137-0277},
Krzysztof~Barczynski\inst{3,4}\orcidlink{0000-0001-7090-6180},
Gabriel~Pelouze\inst{6},
 \and
 Philip~J.~Smith\inst{8}
}

\institute{Max Planck Institute for Solar System Research, Justus-von-Liebig-Weg 3, 37077, G{\"o}ttingen, Germany \\
\email{smandal.solar@gmail.com}
\and
School of Space Research, Kyung Hee University, Yongin, Gyeonggi 446-701, Republic of Korea
\and
Physikalisch-Meteorologisches Observatorium Davos, World Radiation Center, 7260 Davos Dorf, Switzerland 
\and
ETH-Z\"{u}rich, Wolfgang-Pauli-Str. 27, 8093 Z\"{u}rich, Switzerland
\and
Solar-Terrestrial Centre of Excellence -- SIDC, Royal Observatory of Belgium, Ringlaan -3- Av. Circulaire, 1180 Brussels, Belgium
\and
Université Paris-Saclay, CNRS,  Institut d'Astrophysique Spatiale, 91405, Orsay, France
\and
Skobeltsyn Institute of Nuclear Physics, Moscow State University, 119991 Moscow, Russia
\and
UCL-Mullard Space Science Laboratory, Holmbury St. Mary, Dorking, Surrey RH5 6NT, UK
\and
Bay Area Environmental Research Institute, Moffett Field, CA, US
\and
Astrophysics Research Centre, School of Mathematics and Physics, Queen’s University Belfast, University Road, Belfast, BT7 1NN, Northern Ireland, UK
}

%^^^^^^^^^^^^^^^-Abstract-^^^^^^^^^^^^^^^^^^^^
\abstract{
Dynamic fibrils (DFs) are commonly observed chromospheric features in solar active regions. Recent observations from the Extreme Ultraviolet Imager (EUI) aboard the Solar Orbiter have revealed unambiguous signatures of DFs at the coronal base, in extreme ultraviolet (EUV) emission. However, it remains unclear if the DFs detected in the EUV are linked to their chromospheric counterparts. Simultaneous detection of DFs from chromospheric to coronal temperatures could provide important information on their thermal structuring and evolution through the solar atmosphere. In this paper, we address this question by using coordinated EUV observations from the Atmospheric Imaging Assembly (AIA), Interface Region Imaging Spectrograph (IRIS), and EUI to establish a one-to-one correspondence between chromospheric and transition region DFs (observed by IRIS) with their coronal counterparts (observed by EUI and AIA). Our analysis confirms a close correspondence between DFs observed at different atmospheric layers, and reveals that DFs can reach temperatures of about 1.5 million Kelvin, typical of the coronal base in active regions. Furthermore, intensity evolution of these DFs, as measured by tracking them over time, reveals a shock-driven scenario in which plasma piles up near the tips of these DFs and, subsequently, these tips appear as bright blobs in coronal images. These findings provide  information on the thermal structuring of DFs and their evolution and impact through the solar atmosphere.}

   \titlerunning{Evolution of dynamic fibrils}
   \authorrunning{Sudip Mandal et al.}
   \maketitle

%*****************************************
\section{Introduction}\label{sec:intro}
Dynamic fibrils (DFs), one of the prominent chromospheric features of solar active regions, are characterised by their dark, elongated, jet-like appearance in the wings and core of H$\alpha$ \citep{2007ASPC..368...27R}. DFs are thought to be closely related to the quiet-Sun type-I spicules and, likewise, are shock-driven phenomena \citep{2004Natur.430..536D,2006ApJ...647L..73H}. Considering their ubiquitous presence, it is therefore natural to ask whether hotter counterparts of these DFs exist at coronal temperatures. So far, reports of such signatures at coronal temperatures are very few. For example, \citet{2016ApJ...817..124S} reported bright rim-like parabolic structures (indicative of DFs) in space-time plots derived using coronal data from the Atmospheric Imaging Assembly \cite[AIA;][]{2012SoPh..275...17L}. However, they could not reliably identify the bright features that produced those parabolic traces in the first place, primarily due to the inadequate spatial resolution of AIA (image scale of $\sim$430\,km\,pixel$^{-1}$). The 174\,\AA\ High-Resolution Imager of the Extreme Ultraviolet Imager \cite[EUI;][]{2020A&A...642A...8R} on Solar Orbiter \citep{2020A&A...642A...1M} has overcome this limitation by providing high-resolution, high-cadence extreme ultraviolet (EUV) observations. Using EUI data (with an image scale of $\sim$135\,km\,pixel$^{-1}$), \citet{2023A&A...670L...3M} reported the first unambiguous detection of DFs at the coronal base of an active region. Small bright blobs of sizes $\sim$0.5 Mm$^2$ within the EUV moss features of that active region, were found to be moving back and forth with time, producing parabolic tracks in space-time maps. Properties of these blobs  matched well with earlier studies of chromospheric DFs  \citep[e.g.,][]{2007ApJ...655..624D}. Therefore, \citet{2023A&A...670L...3M} hypothesised that the observed bright blob-like features were the hot tips of cooler chromospheric DFs.

%-------------------
\begin{figure*}[!ht]
\centering
\includegraphics[width=0.80\textwidth,clip,trim=0cm 0cm 0cm 0cm]{  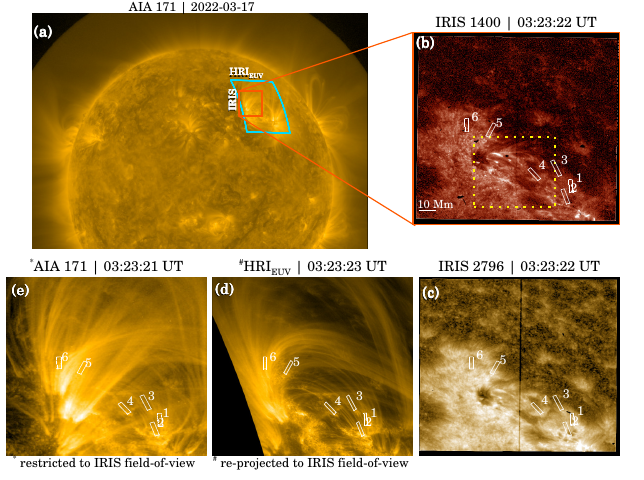}
\caption{Overview of the coordinated EUI-AIA-IRIS observation on 2022-03-17. Panels a-e are ordered clockwise. Panel-a shows the AIA 171~{\AA} image in the background while the IRIS and EUI fields-of-view are marked by red and cyan rectangles, respectively. Panels-b and c present the IRIS 1400~{\AA} and 2796~{\AA} slit jaw images, while panel-d shows the EUI image after reprojecting it to the IRIS field-of-view. Panel-e shows the same but for AIA 171~{\AA} channel (without reprojection). The white boxes on panels-b to e, mark the locations of the artificial slits that are used to derive the space-time maps. The yellow box on panel-b outlines the region analysed in Fig.~\ref{fig:align}.
}
\label{fig:context}
\end{figure*}
%------------------

Nevertheless, the question of whether the bright blob-like features reported in \citet{2023A&A...670L...3M} are of coronal origin ($\log$ T= 6) or of a  cooler transition region plasma ($\log$ T= 5.4), remained open. This is largely because a) the response functions of these coronal imagers span a wide range of temperature, and they often have a secondary peak at lower temperature alongside the high-temperature primary peak and, b) there were no lower temperature diagnostics available for the EUI dataset used by \citet{2023A&A...670L...3M} and therefore, no possibility of independent verification of the temperature structure of DFs. In this work we used co-ordinated EUI, AIA, and IRIS (Interface Region Imaging Spectograph \citealp[IRIS;][]{2014SoPh..289.2733D}) observations and followed the evolution of DFs from the chromosphere to the lower corona. Our results hint towards a more comprehensive understanding of a DF's evolution than we have thus far and also provide insights about their thermal structuring.

%====================================================
\section{Data} \label{sec:data}

The EUV dataset was taken by the 174~{\AA} High Resolution Imager (\hrieuv) of EUI on 2022-03-17, between 03:23:08 UT and 04:08:05 UT, with a cadence of 3~s \citep[part of the SolO/EUI Data Release 5.0;][]{euidatarelease5}. At the time of this observation, Solar Orbiter was located at a distance of 0.379 AU from the Sun, and therefore, one \hrieuv pixel corresponds to 135\,km on the solar surface. Moreover, the angle between the Sun-Solar Orbiter line and the Sun-Earth line was $\approx$26.4$\degree$. We complemented the EUI observation with co-temporal EUV data from AIA, onboard the Solar Dynamics Observatory \cite[SDO;][]{2012SoPh..275....3P}. In particular, we use data from the AIA 171~{\AA} channel, which samples the plasma of a similar temperature to \hrieuv, but with a significantly lower spatial (431 km/pixel) and temporal (12~s) resolution. To capture the lower temperature dynamics, we used coordinated slit-jaw imager (SJI) data from two channels of IRIS, namely the 1400 and 2796~{\AA} channels. These IRIS datasets have a cadence of 3.6~s and a pixel scale of 239 km (owing to its 2x2 spatial binning). Lastly, we correct for the difference in light travel time between Sun-Solar Orbiter and Sun-SDO, and all the time stamps quoted in this paper are Earth times.

%-------------------
\begin{figure*}
\centering
\includegraphics[width=0.95\textwidth,clip,trim=0cm 0cm 0cm 0cm]{  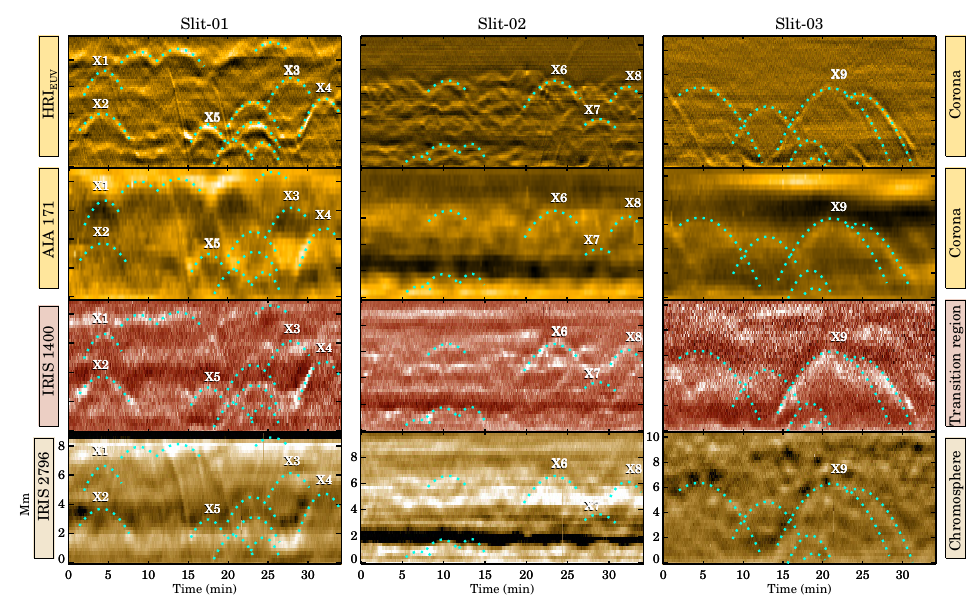}
\caption{Evolution of dynamic fibrils. Space-time (X-T) maps for slit-1 (left column), slit-2 (middle column) and slit-3 (right column) are displayed. The top X-T map in each column is from the \hrieuv data sequence, followed by the maps from AIA 171~{\AA}, IRIS 1400~{\AA} and IRIS 2796~{\AA} data. In each column, the cyan curves outline the parabolic fit to the bright tracks, as seen in the corresponding  EUI X-T map. Tracks marked with letter `\texttt{X}' are discussed further in Section~\ref{sec:xt}.}
\label{fig:xt1}
\end{figure*}
%------------------

 Figure~\ref{fig:context}a shows an AIA 171~{\AA} image from this observational campaign alongside the field of views (FOVs) of  \hrieuv and IRIS, which are outlined via the cyan and red curves. The IRIS observations cover a part of the \hrieuv FOV, and considering our aim of following the evolution of a DF simultaneously along different heights in the solar atmosphere, we restrict the AIA and \hrieuv FOVs to match the FOV of IRIS. Additionally, we re-projected the EUI data onto the IRIS field of view to take into account the angle of 26.4$\degree$ between the Sun-Solar Orbiter and the Sun-Earth line (therefore with AIA and IRIS). This re-projection was carried out using the WCS keywords (present in the data files) as outlined in \citet{2006A&A...449..791T}. The \hrieuv image in Fig.~\ref{fig:context}d shows this re-projection.
 Moreover, as this reprojection relies on a photospheric radius, it may encounter line-of-sight issues for features at higher altitudes. However, low-lying features such as dynamic fibrils, should be least affected.

%++++++++++++++++++++++++++++++++++++++++++++
\section{Analysis and results}

The coordinated EUI-AIA-IRIS observations were centred on the active region NOAA 12965, which was in its decaying phase \citep{2023arXiv230105616B}. A variety of features such as coronal loops and low-lying filaments (in AIA-171 and \hrieuv), spots and moss (in IRIS-1400~{\AA} and 2796~{\AA}) are seen within the FOV.

\subsection{Space-time (X-T) maps}\label{sec:xt}
\begin{figure*}
\centering
\includegraphics[width=0.95\textwidth,clip,trim=0cm 0cm 0cm 0cm]{  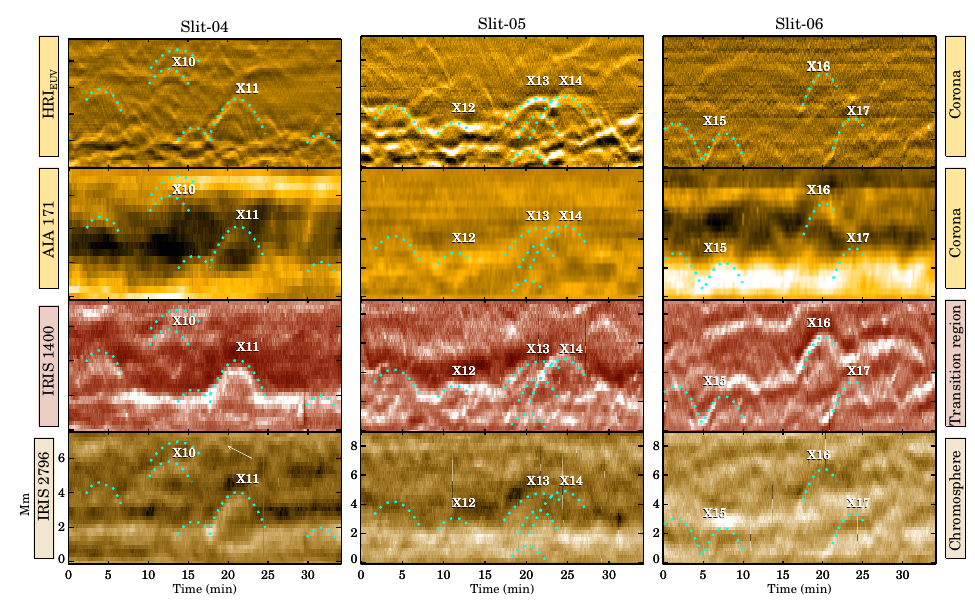}
\caption{Same as Fig.~\ref{fig:xt1} but for slit-4 (left column), slit-5 (middle column) and slit-6 (right column). The arrow in the 2796 map of slit-4 points to a particular case discussed in Sect.~\ref{sec:xt}.}
\label{fig:xt2}
\end{figure*}

To capture DFs and their evolution, we placed multiple artificial slits in and around the active regions as shown in Fig.~\ref{fig:context}. The slit positions were fixed after visually inspecting different locations within the FOV and then picking out the ones from which we have good signal in the IRIS channels. Each slit is 10 IRIS pixels wide ($\sim$2.4 Mm) and the final X-T maps are derived after averaging emission over the respective slit widths. Furthermore, to enhance the appearance of bright structures, we subtracted the background (as calculated through a boxcar smoothing) along the transverse direction (i.e., along the y-axis) of an X-T map. Figures-\ref{fig:xt1} and \ref{fig:xt2} present these contrast-enhanced X-T maps from six of these slits. Additionally, these maps with individual colorbars can be found in Appendix.~\ref{sec:xt_colorbar}.

%-------------------
%------------------

 The \hrieuv X-T maps in Fig.~\ref{fig:xt1} and Fig.~\ref{fig:xt2} are filled with bright, seemingly parabolic tracks that are basically EUV signatures of DFs \citep{2023A&A...670L...3M}. Sometimes these tracks appear repetitive, highlighting the recurrent nature of DFs (for example, \hrieuv Slit-1 X-T map between y=7 and 8 Mm). Since a given DF is associated with a given parabolic profile, it is difficult to say from these images whether the repetitive tracks were due to the same DF or different DFs that happen to be in close spatial proximity. Nonetheless, we fitted few of these \hrieuv tracks\footnote{We visually identified and selected these 41 tracks as they appear relatively prominent, in parts or as a whole. These selected set of tracks, as representative examples, sample  each of the DF category. Figures~\ref{fig:xt1} and ~\ref{fig:xt2} contain all 41 of these detected tracks}. with parabolas (see \citealp{2023A&A...670L...3M} for details of this fitting procedure) and the fitted curves are overplotted in cyan in each of these \hrieuv maps. For a given slit, we then overplotted these fitted curves on all other X-T maps i.e., the cyan curves in AIA and IRIS X-T maps are the ones we fitted in the corresponding \hrieuv X-T maps. A closer look at each of these X-T maps, immediately reveals that the visibility of these parabolic trajectories of the DFs is best in \hrieuv (which has the highest spatial resolution), whereas for AIA-171, the trajectories are either less resolved and fainter or absent. For IRIS- 1400 channel, the signal is {more prominent}, while for 2796 channel, it is mostly at the base of these trajectories. Depending upon their simultaneous appearance in different channels, we classified the observed DFs into the following four categories:\\
 
{\it Category-I; Visible only in \hrieuv}. An example of this is the curve X7 in Fig.\ref{fig:xt1}. At first, it appears that DFs that have a small maximum height, i.e. those that travel only a few Mm into the atmosphere, are the ones that fall into this category. However, examples such as X1 and X3 which have average maximum heights but still lack signals in other channels, are exceptions to this narrative.

{\it Category-II; Visible only in \hrieuv, AIA 171 and IRIS 1400 maps:} Examples of such DFs are X5 and X8 in Fig.~\ref{fig:xt1}b, and X12 and X15 in Fig.~\ref{fig:xt2}. There is, however, diversity within this class. For example, in some cases, the bright track in IRIS-1400, is significantly brighter either at the beginning or at the end of the track (e.g., X8 or X15), while in other cases, it is uniformly bright throughout its whole extent (e.g., X5 or X12)\footnote{This is not an artefact of the contrast enhancement procedure as the same trend is also visible in the original maps.}. 

{\it Category-III; Visible in all four channels:} Although for these DFs we find a simultaneous signal in X-T maps of all four channels, their signatures in the IRIS 2796\,\AA\ channel (and seldom in IRIS 1400\,\AA\ channel) are either often faint or only localised to the footpoints of the tracks. There could be several factors affecting the appearance of a given DF in any IRIS passband, including spectarl and thermal characteristics of the filters. However, investigation of these factors is beyond the scope of this paper. Nonetheless, the curve X9 in Fig.~\ref{fig:xt1} and X11 in Fig.~\ref{fig:xt2} are among the best examples of this category. The other ones such as X4, X6 of Fig.~\ref{fig:xt1}  and X10, X16, X17 of Fig.~\ref{fig:xt2} also fit in here despite their ambiguous appearances in the IRIS 2796\,\AA\ maps.

{\it Category-IV; Exceptions:} This category contains examples that we could not fit into any of the previous three categories. For example, X2 in Fig.~\ref{fig:xt1} has a clear visibility in \hrieuv and 1400 channels (with some hints in the IRIS 2796 maps), but the signal in AIA 171 channel is (almost) non-existent. The other extreme example is the case in Fig.~\ref{fig:xt2}, where we see a (relatively) clear parabolic signal in 2796 channel (next to X10, between t=16 and 20 min and y=6 and 7 Mm, as highlighted by the arrow), while its signature in other channels is almost non-existent.

%-------------------
\begin{figure*}
\centering
\includegraphics[width=0.85\textwidth,clip,trim=0cm 0cm 0cm 0cm]{  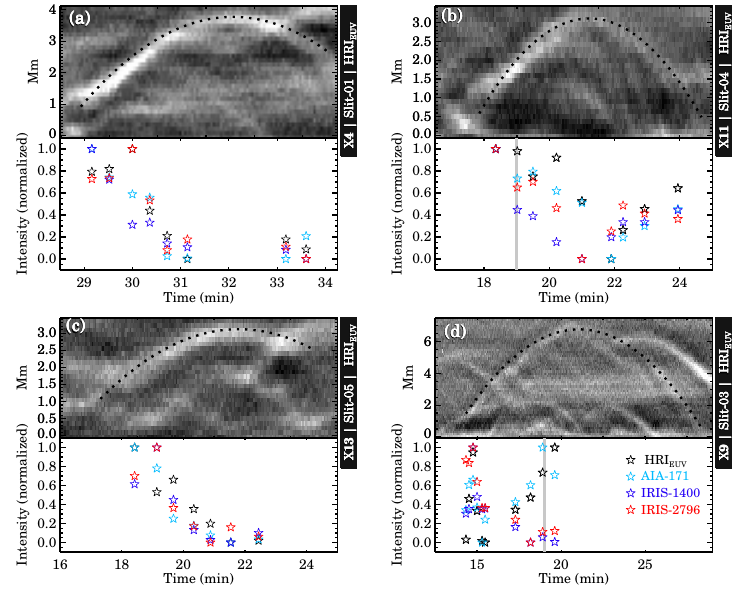}
\caption{Intensity evolution of four selected DFs. 
In Panel-a, the top section shows the \hrieuv X-T maps (between t=28 and 34 min) of slit-1, while the bottom section shows the evolution of intensities in four different imaging channels (\hrieuv in black, AIA 171~{\AA} in cyan, IRIS 1400~{\AA} in blue and IRIS 2796~{\AA} in red), of the DF that created the parabolic trajectory in the X-T map. The black dotted line in the top section is the parabolic fit to the observed bright track. Panels-b, c and d show the same but for slit-4, slit-5 and slit-3. The grey vertical lines in panels-b and d indicate the time when the emission measure analyses as shown in Fig.~\ref{fig:dem} and ~\ref{fig:dem2} are performed. Animations associated to this figure, are available \href{https://drive.google.com/drive/folders/17-fqQz_P2T18llJ1jB6MJISMRvT5063F?usp=sharing}{here}.
}
\label{fig:lc}
\end{figure*}
%------------------

\subsection{Temporal evolution of DF intensities}
Having demonstrated that some DFs show signatures in different temperature regimes of the solar atmosphere, we will now investigate their evolution by tracking their emission over time. Although one can track a DF's temporal evolution by following the bright track that it creates in an X-T map, we choose to do it by following a DF in every frame in a data sequence. The issue with first approach has to do with the fact that the X-T maps, as shown in Fig.~\ref{fig:xt1} and \ref{fig:xt2}, were produced after averaging over their slit widths and, therefore, other features encroaching that slit will also contribute to the derived intensity\footnote{It is practically not possible to adjust the slit width so that it only follows a single DF over all frames.}. Furthermore, it is also affected by the significant amount of overlap from other DFs as well. These issues, however, can be mitigated through latter method where one follows a DF in every frame avoiding overlaps. In order to make sure that we are following the same pixels in all four channels, we re-scale the \hrieuv and AIA data to match the IRIS resolution (effectively, this means up-scaling the AIA data and down-scaling the \hrieuv data). Such re-scaling provides an added advantage that any pixel level mismatch in DF locations between \hrieuv and IRIS (whether physical or due to re-projection of the \hrieuv data) would also be taken care of in this process. Given its better signal-to-noise ratio compared to other channels, we performed the feature tracking on the \hrieuv images. Finally, the tracking was done visually, i.e., by going through frame-by-frame and using the cursor to select the centre of a bright blob. We calculated the intensity of the detected feature as the mean of all the pixels within a circular region encompassing the full extent of the blob (see the animation associated with Fig.~\ref{fig:lc} for more details). Once a DF has been traced in \hrieuv, we used the same positional information to extract intensity values from the remaining three channels.

Figure~\ref{fig:lc} presents the intensity evolution of four selected DFs (three DFs from category-III (panel-a,b,d) and one from category-II (panel-c)). We find a common trend for cases shown in panels-a, b and c. DF intensities in all four channels systematically decrease as it reaches its maximum height in its corresponding X-T map. However, once a DF starts receding (i.e., during the descending part of a parabola), there is a hint of intensity enhancements in all four channels. Nonetheless, the situation is quite different for the case shown in Fig.~\ref{fig:lc}d, where we instead find a rapid drop in intensity at the beginning of the ascending phase (in line with the previous three cases). However, the situation changes quickly as the DF approaches its maximum height, where we see a systematic increase in \hrieuv and AIA-171 intensities whereas the IRIS-1400, 2796 intensities drop simultaneously. Unfortunately, we could not trace this DF beyond its maximum height due to substantial contribution from background (or foreground) features (see the associated movie). We do not include error estimates due to the photon Poisson (shot) noise, electronic readout noise, compression noise, dark current noise associated to the derived intensities (statistical errors are not very useful here as the sample size is small in each case). Nevertheless, the commonality and concurrency of the observed trends in intensities derived from data of different passbands of different instruments, weigh in favour of the observed signal being of solar origin.

%=+=+=+=+=+=+=+=+=+=+=+=+=+=+=+=+=+=+=
\section{Discussion and conclusion}
%------------------
\begin{figure*}
\centering
\includegraphics[width=0.80\textwidth,clip,trim=0cm 0cm 0cm 0cm]{  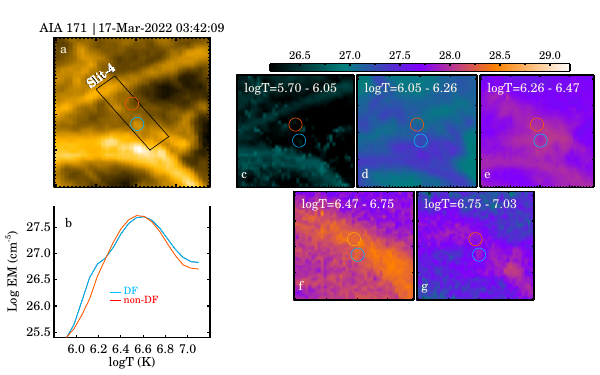}
\caption{Emission measure (EM) analysis of a DF. Panel-a shows an AIA 171~{\AA} image with the DF outlined by the cyan circle. The EM curve (derived at 03:42 UT) of the central pixel of that circle is shown in panel-b (the blue curve) while the red curve shows the same but for a pixel away from the DF as outlined by the red circle in panel-a.  Panels-c to g show 2D EM maps (of the FOV shown in panel-a) at specific temperature bins as mentioned on each panel. 
}
\label{fig:dem}
\end{figure*}
%------------------

In this work, we set out to explore the connection between the lower-temperature chromospheric DFs with that of EUV DFs through coordinated \hrieuv, AIA and IRIS observations. Our results indicate strong correlations between them, both spatially and temporally. Below we highlight our main findings:
\begin{itemize}
    \item The EUI X-T maps derived from moss-type regions are found to be filled with bright parabolic tracks which are indicative of DFs. Corresponding AIA X-T maps are also similarly populated while signatures of such tracks are less frequent in IRIS 1400 \AA\, and 2796 \AA\, data. We also note that the signal in 2796 \AA\ channel is stronger near the start of these parabolic tracks, and some of the tracks in 1400 \AA\ channel show a hint of spatial offset with the same in \hrieuv data.\\

    \item There exist several cases where we found simultaneous signal in \hrieuv, AIA 171~{\AA} and IRIS 1400~{\AA} channels. However, DF events where all four channels show co-spatial and co-temporal signals are infrequent.\\
    
    \item By following the evolution of four selected DFs, we
    found that the intensity of a DF tends to decrease as it travels upward in the atmosphere and, once the DF crosses its maximum height, its intensity starts to increase again. We, however, also found an example which is an exception to this scenario, therefore demonstrating the need for more statistics.  \\
    
\end{itemize}

As mentioned in the introduction, DFs have traditionally been identified as a shock-driven phenomenon \citep{2005ApJ...624L..61D,2006ApJ...647L..73H}. In this scenario, magnetoacoustic waves from the lower atmosphere steepen into shocks as they propagate upward and push the chromospheric material to higher heights to form a DF. If we now consider the bright blobs in \hrieuv images as the hotter counterparts of the same chromospheric DFs, then one would expect to see a definite signature in IRIS channels for each bright track seen in \hrieuv (or AIA) images. This is, however, not the case, as we found in Fig.~\ref{fig:xt1} and Fig.~\ref{fig:xt2}. There could be several reasons for the absence of such one-to-one correspondence: (1) A DF in the IRIS-2796 channel appears as more of an elongated diffuse structure as opposed to a bright blob that we found in other channels. Therefore, their visibility in an X-T map improves only if the DF is significantly brighter than the background. (2) Secondly, the properties of IRIS slit-jaw passbands also influence the visibility of these DFs. For example, the IRIS 1400~{\AA} slit jaw imager passband is relatively wide (55 \AA), and in non-flaring conditions, it is the Si I recombination continuum that contributes most to this channel \citep{2015ApJ...803...44M}. On the other hand, the 2796~{\AA} bandpass is significantly narrower (5~\AA), but the Mg II k line line is formed over a considerable range of heights and with strong non-local thermodynamic equilibrium conditions \citep{2013ApJ...772...90L}. Therefore, DFs with lower contrast are barely detectable while the brightest ones remained visible. Furthermore, the IRIS dataset we used in this study has a 2$\times$2 binning which results in a pixel scale of 0.33\arcsec. Thus, the spatial binning of this IRIS data may also have played a role in the poor visibility of DFs. (3) Lastly, we recall that there exists a moderate angle of 26.4$\degree$ between the two spacecraft (IRIS and Solar Orbiter). Hence, different alignment of (magnetic) structures may have influenced their visibility across instruments.

Clues about the mechanism that possibly makes a chromospheric DF visible to higher temperature channels are found in Fig.~\ref{fig:lc}. Those three cases where we see simultaneous decrement in intensity in all four channels can be explained through the following evolutionary scenario: The chromospheric material, propelled by the shock, travels upward and piles up material near the tip of the DF (similar to Fig-5 of \citealp{2016ApJ...829L..18B}). This material pile-up enhances their visibility in the \hrieuv and AIA 171~{\AA} channels and is also the reason behind their blob-like appearance in these channels. As the DF moves upward in the atmosphere, it constantly loses energy, primarily through radiation, with contribution from thermal conduction and geometrical damping (fanning out of the magnetic field with height). Therefore, it explains why all four channels show a simultaneous intensity decrease. However, the reason why the intensity starts to increase as the DF starts to recede is not yet apparent. One possibility could be that the falling material gets adiabatically compressed against the denser chromospheric plasma, leading to a temperature increase in turn causing the emission enhancement. We cannot, however, rule out another possibility in which the increase in intensity is simply due to enhanced density along the LOS. This evolutionary scenario matches well with the type-I spicules \citep{1968SoPh....3..367B}.

However, the case shown in Fig.~\ref{fig:lc}d, has a different evolution. Here, we postulate that the scenario in which the intensity of cooler channels drops and simultaneously the intensity of the hotter channels increases is similar to a Type-II spicule \citep{2009ApJ...701L...1D}. Spicules in general, undergo complex evolution across different atmospheric layers \citep{2012ApJ...759...18P}, while the Type-II ones often leave their imprints in transition region and coronal images \citep{2015ApJ...799L...3R, 2019Sci...366..890S,2021A&A...647A.147B}. Numerical modeling works such as by \citet{2018ApJ...860..116M} reveal that these features are associated with magneto-acoustic shocks and flows, and also supply mass to coronal loops.
Nonetheless, Type-II are a subclass of spicules that reach greater heights and move faster. Furthermore, from the figure (Fig.~\ref{fig:lc}), it is also evident that this particular DF shows a significantly larger height parameter ($\approx$6 Mm) compared to other examples ($\approx$3 Mm)\footnote{Although 6 Mm is longer than expected for a DF, it is unlikely to be another type of structure, such as a small filament. However, a more detailed morphological comparison is outside the scope of our study.}. Therefore, this may be an example of a EUV counterpart of a Type-II spicule. However, at the same time, we are cautious about this conclusion because a) it is based on a case which has a complicated evolutionary track with many overlapping structures (see the animation associated with Fig.~\ref{fig:lc}) and b) this DF could well be travelling nearly parallel to the solar surface such that higher height does not always mean higher altitude. Therefore, this needs further investigation. Future coordinated observations including ground-based telescopes (specifically observations in a passband centred on H$\alpha$) would help in better understanding the lower atmospheric evolution of these DFs.

 Lastly, we discuss the possibility of a DF reaching typical coronal temperatures ($\sim$ MK). To this end, we calculated the emission measure (EM) of DF plasma using the co-spatial and near-simultaneous multi-wavelength EUV data from AIA. We followed the inversion technique of \citet{2015ApJ...807..143C}. Figure~\ref{fig:dem} presents one such case (another case is shown in the appendix). The EM curve of the DF (outlined by the blue circle in panel-a) is similar to a typical active region EM, and it has a peak emission at $\log$T of 6.6 (panel-b) while a secondary hump is seen at $\log$T=6.2. For comparison, we overplot (in red) the EM curve from a location away from the DF, as outlined by the red circle. Interestingly, the peak of the curve remains at $\log$T=6.6, while the secondary hump seems to be absent in this case. However, we cannot draw a definitive conclusion on the exact temperature of the DF without further investigation of the DEMs by tracking the DF over time, which is beyond the scope of this study. Nonetheless, at first glance, it appears that the DF is indeed reaching a temperature higher than 1 MK ($\approx$ 1.5 MK). We further generated multiple 2D EM maps by dividing the entire temperature range into several bins (panels-c to g). Through these maps, we found signatures of a hot loop nearly in the line-of-sight of the DF in question (panel-f) and could probably be responsible for the observed primary peak at a higher temperature, while the DF itself might have a slightly lower temperature, but still slightly over 1\,MK. Therefore, at least some DFs can well be considered as a source of coronal emission in active regions. Furthermore, recent studies on spicules,  rapid blueshifted excursions (RBEs), and rapid redshifted excursions (RREs) also suggest that some of these features indeed show signatures in transition region and coronal observations \citep{2021A&A...654A..51B,2022MNRAS.515.2672V,2023MNRAS.524.1156V}. As discussed earlier, since DFs are closely related to spicules, these studies align well with the results we find here (although some of these upper atmospheric signatures could also be due to type-II class spicules, e.g., as shown in \citealp{2019Sci...366..890S} and therefore, may not be related to DFs).

To conclude, by analysing coordinated \hrieuv-IRIS-AIA observations of a moss-type region, we found a clear association between DFs that appear blob-like in \hrieuv and IRIS-1400 data, with the DFs in 2796 data that appear more as elongated, diffuse features. These DFs have a temperature of $\approx$ 1.5 MK, i.e. typical coronal values. Temporal analysis of their intensity evolution revealed a scenario that is similar to type-I spicules.

\begin{acknowledgements}
We thank the anonymous reviewer for the encouraging comments and helpful suggestions. Solar Orbiter is a space mission of international collaboration between ESA and NASA, operated by ESA. The EUI instrument was built by CSL, IAS, MPS, MSSL/UCL, PMOD/WRC, ROB, LCF/IO with funding from the Belgian Federal Science Policy Office (BELSPO/PRODEX PEA 4000112292 and 4000134088); the Centre National d’Etudes Spatiales (CNES); the UK Space Agency (UKSA); the Bundesministerium für Wirtschaft und Energie (BMWi) through the Deutsches Zentrum für Luft- und Raumfahrt (DLR); and the Swiss Space Office (SSO). We are grateful to the ESA SOC and MOC teams for their support. Solar Dynamics Observatory (SDO) is the first mission to be launched for NASA's Living With a Star (LWS) Program. The data from the SDO/AIA consortium are provided by the Joint Science Operations Center (JSOC) Science Data Processing at Stanford University. IRIS is a NASA small explorer mission developed and operated by LMSAL with mission operations executed at NASA Ames Research Center and major contributions to downlink communications funded by ESA and the Norwegian Space Centre. L.P.C. gratefully acknowledges funding by the European Union (ERC, ORIGIN, 101039844). Views and opinions expressed are however those of the author(s) only and do not necessarily reflect those of the European Union or the European Research Council. Neither the European Union nor the granting authority can be held responsible for them. DML is grateful to the Science Technology and Facilities Council for the award of an Ernest Rutherford Fellowship (ST/R003246/1). ANZ and LR thank the Belgian Federal Science Policy Office (BELSPO) for the provision of financial support in the framework of the PRODEX Programme of the European Space Agency (ESA) under contract numbers 4000136424 and 4000134474. JMS gratefully acknowledges support by NASA grants 80NSSC18K1285, 80NSSC20K1272, 80NSSC21K0737, 80NSSC21K1684, 80NSSC23K0093 and contract NNG09FA40C (IRIS) and NNH19ZDA013O (MUSE). 
\end{acknowledgements}

\begin{appendix}

\section{Further examples of EM analysis}
\begin{figure*}
\centering
\includegraphics[width=0.85\textwidth,clip,trim=0cm 0cm 0cm 0cm]{  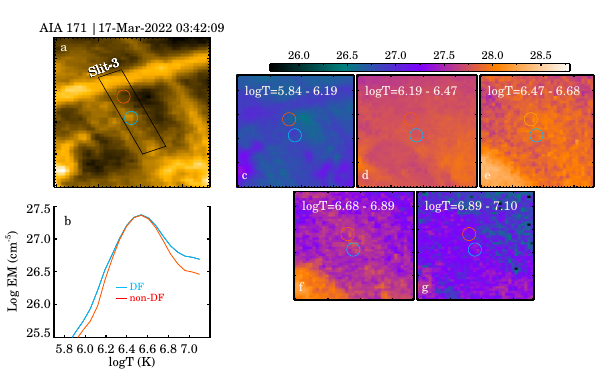}
\caption{Same as Fig.~\ref{fig:dem} but for a different DF (from slit-3). 
}
\label{fig:dem2}
\end{figure*}

\begin{figure*}
\centering
\includegraphics[width=0.85\textwidth,clip,trim=0cm 0cm 0cm 0cm]{  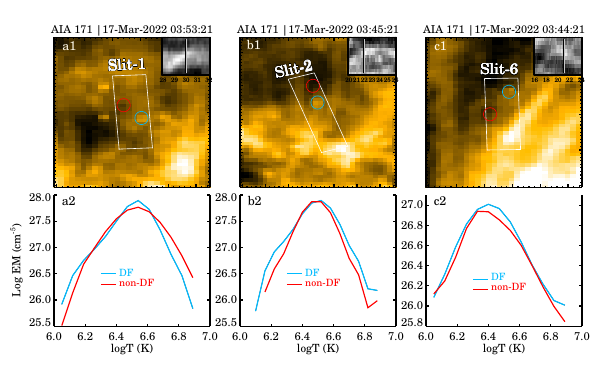}
\caption{Additional examples of EM analysis of DFs. Panel descriptions are same as that of Fig.~\ref{fig:dem}. The tracks that those DFs create in AIA 171 maps are shown in insets of panel-a1, b1 and c1. The vertical lines in these inset panels highlight when the EM measurement was carried out.}
\label{fig:dem_stat}
\end{figure*}

In Fig.~\ref{fig:dem2} we present an additional example of the EM analysis on a DF. This particular DF is located within slit-3. The EM curve in this example (panel-b) peaks at $\log\,T=6.6$ and appears similar to the curve presented in Fig.~\ref{fig:dem}b, although the secondary hump at $\log\,T=1.2$ is not that pronounced as before. Unlike the previous case, the temperature binned 2D EM maps (panels-c to g) do not show any blob-like structure but rather a diffuse emission near the peak temperature, covering the whole field of view. We present further examples of EM analysis in Fig.~\ref{fig:dem_stat} which contains examples from slit-1, 2 and 6. Through these, we find that some DFs show signatures of a secondary peak around $\log\,T=1.2$ (e.g., slit-2) while others do not (e.g., slit-6). Therefore, it further emphasises the need for a detailed statistical investigation exploring the contribution of the foreground and background structures to the derived EM curve (and its time evolution).

%------------------
%------------------
%------------------
%------------------

\section{Correlation analysis}

\begin{figure*}
\centering
\includegraphics[width=0.85\textwidth,clip,trim=0cm 0cm 0cm 0cm]{  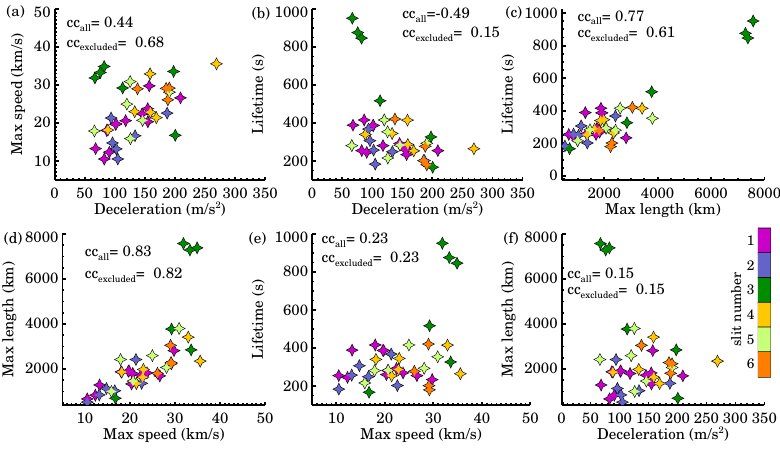}
\caption{Inter-relationships between different DF parameters. Data from different slits are represented by a color scheme as mentioned in panel-f. The correlation coefficients calculated with all data points (cc$_{all}$) and excluding three outlier points (cc$_{excluded}$) are also printed on each panel. See text for more details.}
\label{fig:cor}
\end{figure*}

Inter-relationships between DF parameters (such as lifetime, apparent speed, travelled length, etc.) contain information about their generation mechanism \citep{2007ApJ...655..624D}. Following the methodology outlined in \citet{2023A&A...670L...3M}, we derived the DF length as the maximum height of the parabola and DF lifetime as the time between the two ends of the parabola (after extrapolating the fitted curve to make it symmetric). In Fig.~\ref{fig:cor}, we show such scatter plots for this study. The correlation coefficients we obtained here are similar to those quoted in \citet{2023A&A...670L...3M} and in agreement with the theoretical work by \citet{2007ApJ...666.1277H}. There is, however, more to this story. Three DFs from slit-3 (the dark green symbols) appear as `outliers' in all these scatter plots. A quick look at Fig.~\ref{fig:xt1} reveals these `outlier' DFs are the ones with large tracks and longer lifetimes. Although all parameters of these three DFs are substantially different from other DFs, their maximum speeds seem to be not so extravagantly different. One possible explanation of such behaviour could be related to the inclination of local magnetic field and the periodicity of the driver as found in a numerical simulation by \citet{2007ApJ...666.1277H}. These authors found that although an inclined field hosts a stronger driver (owing to lower acoustic cutoff; \citealp{2004Natur.430..536D,2006ApJ...647L..73H}), it does not necessarily translate into higher maximum speed as this larger driving only leads to larger dissipation. Therefore, these seemingly outlier points that we find in our study, could well be cases with larger inclined field geometry. As we noted earlier, multiple overlappings and atypical evolution of these DFs (X9 from Fig.~\ref{fig:xt1} being one of them) might have an influence on their derived parameters. There is also the complexity of projection effects that can play a role on our results. Therefore, one needs to investigate further, probably with other EUI datasets of equal or better resolution than the one we used in this study.

%------------------
%------------------

\section{Zoomed-in images and Evolution of DFs}
\begin{figure*}
\centering
\includegraphics[width=0.85\textwidth,clip,trim=0cm 0cm 0cm 0cm]{  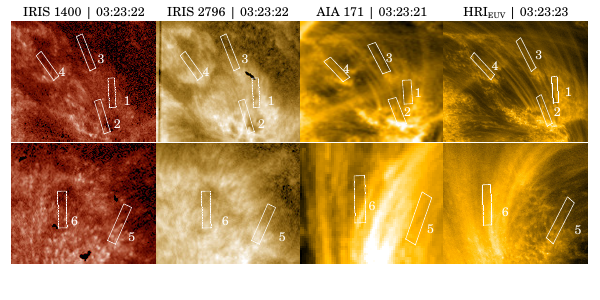}
\caption{A zoomed-in view of Fig.~\ref{fig:context}.}
\label{fig:zoomed_in}
\end{figure*}

We first present a zoomed-in view of the context image (Fig.~\ref{fig:context}) in Fig.~\ref{fig:zoomed_in}. As we see, the artificial slits are placed either parallel or near parallel to the loops that are most likely in the foreground of the DFs. Such placement of slits also makes sure that the trajectories we find in the X-T plots (Fig.~\ref{fig:xt1} and Fig.~\ref{fig:xt2}) are not signatures of kink oscillations \citep{2004SoPh..223...77V,2015A&A...583A.136A,2011ApJ...736L..24O,2022A&A...666L...2M}. This is because, during kink oscillations, a loop sways sideways, and thus, to capture these kink oscillations in an X-T map, we must place the slit perpendicular to the loop, not parallel to it. Hence, the tracks we find in X-T maps are due to motions of blob-like DFs (as discussed below) and not related to transverse oscillations.

%------------------
\begin{figure*}%
    \centering
    \includegraphics[width=0.87\textwidth,clip,trim=0.2cm 0cm 1.1cm 0cm]{  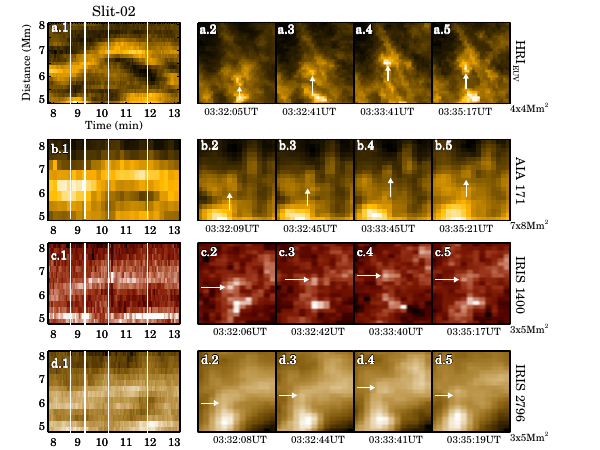}
    \caption{ Evolution of a DF (from slit-02) across channels. Panel a.1 shows the bright (parabolic) track that the DF creates in the EUI X-T map. The four vertical lines mark the time-stamps of snapshots in panels a.2, a.3, a.4, and, a.4, in which the white-arrows point toward the instantaneous position of the DF. Panels b, c and d have the same format but for AIA 171~{\AA}, IRIS 1400~{\AA}, IRIS 2796~{\AA} observations. An animated version of this figure is available online (\href{https://drive.google.com/file/d/18HE0VyGs3ELfx8mZgkE5SScudS0-m_C7/view?usp=drive_link}{here}). }%
    \label{fig:df_evolution1}%
\end{figure*}

\begin{figure*}%
    \centering
    \includegraphics[width=0.87\textwidth,clip,trim=0.2cm 0cm 1.1cm 0cm]{  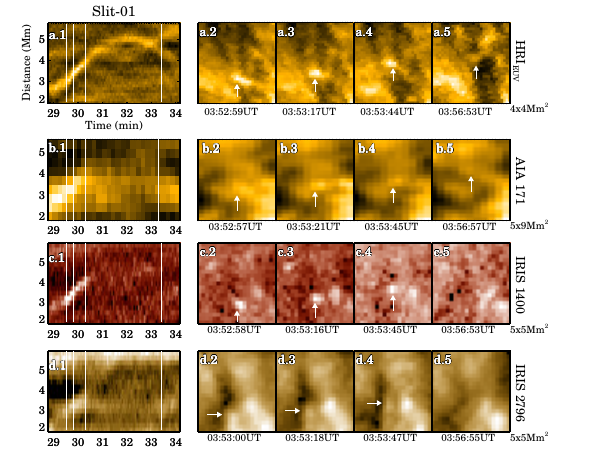}
    \caption{ Similar to Fig but for a DF from slit-01. An animated version of this figure is available online (\href{https://drive.google.com/file/d/1GZmMSenLsV2qhJrJY2qaI-1ybkMd6bm4/view?usp=drive_link}{here}). }%
    \label{fig:df_evolution2}%
\end{figure*}

%------------------
 In Fig.~\ref{fig:df_evolution1} and Fig.~\ref{fig:df_evolution2}, we present two illustrative examples of DF evolution across channels. Furthermore, we also provide animations associated with this figure that shows the complete evolution of these DFs over their lifetime. As seen from the figure, a bright blob-like feature moves back and forth with time and therefore, creating parabolic tracks in X-T maps. The feature appears more diffuse and elongated in IRIS 2796 channel in comparison to other passbands. Moreover, it is best identified in EUI and IRIS 1400 data, while in the AIA 171 data, it can be identified in hindsight, i.e., after spotting it in the EUI images. Therefore, we conclude that although there are co-spatial and co-temporal signatures of DFs from the chromosphere to the corona, their detection is generally limited by the resolution of the data used to investigate them. 

%---------------------------------------

%--------------------------------------

\section{Co-alignment between instruments}

\begin{figure*}
\centering
\includegraphics[width=0.85\textwidth,clip,trim=0cm 0cm 0cm 0cm]{  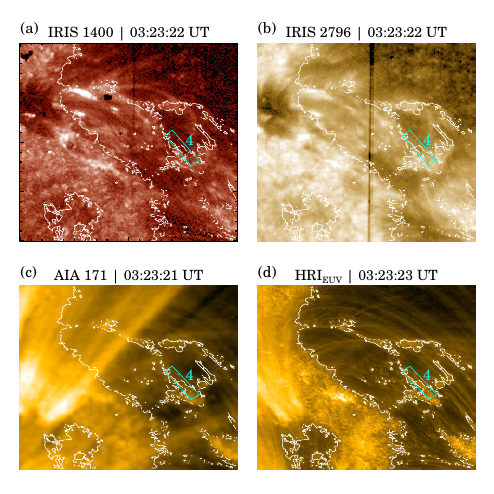}
\caption{Co-aligned datasets from IRIS (panel-a, b), AIA (panel-c) and \hrieuv (panel-d). All four panels of this figure represent a small section of the full field of view as highlighted by the yellow box in Fig.~\ref{fig:context}b. The AIA and \hrieuv images are scaled to IRIS pixel scale. The contours overplotted on every panel are derived using intensities from the \hrieuv image (panel-d). Slit-4 that falls within this FOV is shown as the cyan box in every panel.}
\label{fig:align}
\end{figure*}

As explained in Sect.~\ref{sec:data}, we used WCS keywords to co-align data from \hrieuv, AIA and IRIS. Figure~\ref{fig:align} provides a visual representation of such aligned data. The \hrieuv and AIA images are scaled to the IRIS resolution for easy comparison. Following the contours (derived from the \hrieuv intensity, panel-c), we find that the bright, low-lying regions in all four channels overlap well, while large fan-loops appear somewhat different.    

%------------------
%------------------

\section{X-T maps with colorbars} \label{sec:xt_colorbar}

\begin{figure*}
\centering
\includegraphics[width=0.95\textwidth,clip,trim=0cm 0cm 0cm 0cm]{  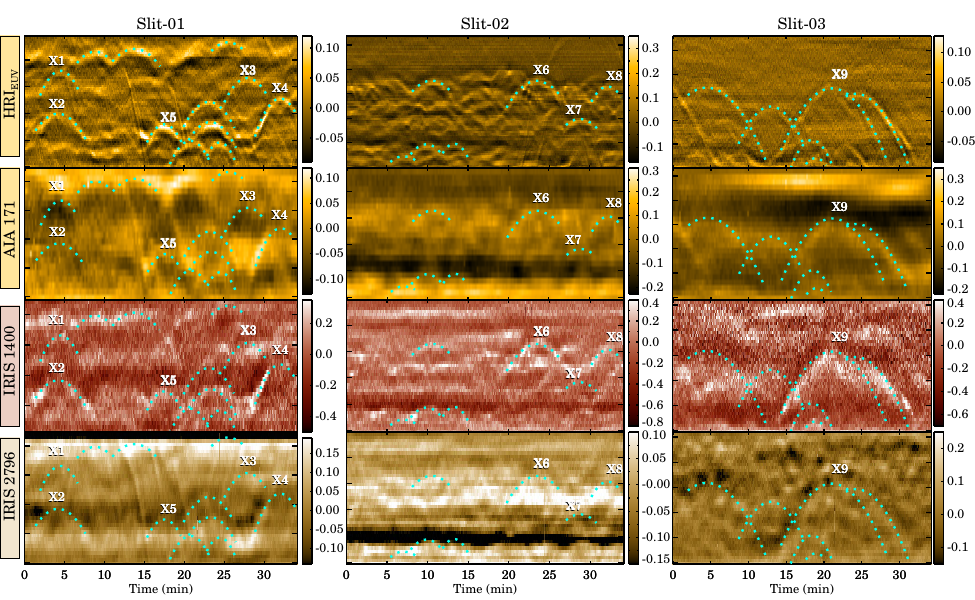}
\caption{Same as Fig.~\ref{fig:xt1} but with colorbars that outline the relative change as described in Sect.~\ref{sec:xt_colorbar}.} 
\label{fig:xt1_colorbar}
\end{figure*}

\begin{figure*}
\centering
\includegraphics[width=0.95\textwidth,clip,trim=0cm 0cm 0cm 0cm]{  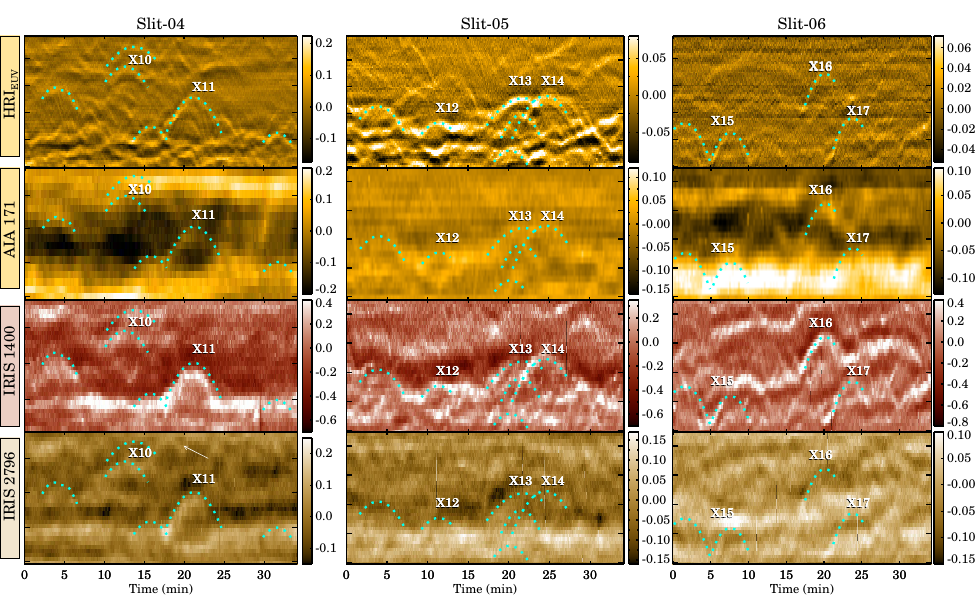}
\caption{Same as Fig.~\ref{fig:xt2} but with colorbars that outline the relative change as described in Sect.~\ref{sec:xt_colorbar}.} 
\label{fig:xt2_colorbar}
\end{figure*}

In Fig.~\ref{fig:xt1_colorbar} and \ref{fig:xt2_colorbar} we present X-T maps (of Fig.~\ref{fig:xt1} and \ref{fig:xt2}) with colorbars for the ease of following the observed intensity changes. Again, we emphasise that these maps are background subtracted (as calculated through a boxcar smoothing along the transverse direction, i.e., along the y-axis of an X-T map). Therefore, the colorbars represent the relative change defined as (map-smoothed map)/smoothed map.

%-------------------
%------------------
%-------------------
%------------------
\end{appendix}

%=+=+=+=+=+=+=+=+=+=+=+=+=+=+=+=+=+=+=
\bibliographystyle{aa}
\bibliography{ref_DF}

\end{document}